\documentclass[aps,prl,twocolumn,showpacs]{revtex4}

\usepackage{amsmath}
\usepackage{amssymb}
\usepackage{graphicx}
\usepackage{dcolumn}
\usepackage{bm}

\newcommand{\eps}{\varepsilon}
\newcommand{\la}{\lambda}

\newcommand{\br}{{\bf r}}
\newcommand{\prt}{\partial}
\newcommand{\bu}{{\bf u}}

\begin{document}

\title{
Oblique  dark solitons in supersonic flow of a Bose-Einstein
condensate}
\author{G.A. El$^{1}$}
\email{G.El@lboro.ac.uk}
\author{A. Gammal$^{2}$}
\email{gammal@if.usp.br}
\author{A.M. Kamchatnov$^{3}$}
\email{kamch@isan.troitsk.ru}

\affiliation{ $^1$ Department of Mathematical Sciences, Loughborough
University, Loughborough LE11 3TU, UK \\
$^2$ Instituto de F\'{\i}sica, Universidade de S\~{a}o Paulo,
05315-970, C.P.66318 S\~{a}o Paulo, Brazil\\
$^3$Institute of Spectroscopy, Russian Academy of Sciences, Troitsk,
Moscow Region, 142190, Russia }

\date{\today}

\begin{abstract}
In the framework of the Gross-Pitaevskii mean field approach it is
shown that the supersonic flow of a Bose-Einstein condensate can
support a new type of pattern---an oblique dark soliton.  The
corresponding exact solution of the Gross-Pitaevskii equation is
obtained. It is demonstrated by numerical simulations that oblique
solitons can be generated by an obstacle inserted into the flow.
\end{abstract}

\pacs{03.75.Kk}

\maketitle

{\it Introduction.} It is known that the nonlinear and dispersive
properties of a Bose-Einstein condensate (BEC) can lead to the
formation of various nonlinear structures (see, e.g.,
\cite{ps2003}).  Until recently, most research has been focused on
experimentally observed vortices and bright and dark solitons.  Furthermore,
the formation of dispersive shock waves in BECs with repulsive
interactions between atoms was considered theoretically in
\cite{kgk04,damski04} and studied experimentally in rotating
\cite{simula} and non-rotating \cite{hoefer} condensate, where it
was shown that dispersive shocks are generated as a result of the
evolution of large disturbances in the BEC.  However, another
important type of nonlinear structure, namely a spatial dark soliton, can
also be realized in a BEC.  The first experimental evidence of their
generation has recently appeared \cite{cornell05}.  In fact,
the existence of oblique spatial solitons in a BEC has a natural
physical basis if the Cherenkov generation of dispersive sound waves
by a small obstacle in the supersonic flow of a BEC is considered and
the effect of increasing the size of the obstacle (i.e.\ the amplitude
of the waves) is determined.  Evidently, along with
dispersion, nonlinear effects become equally important at finite
distances from the obstacle, so that the Cherenkov cone breaks-up
into a spatial structure consisting of one or several dark
solitons.  Such a structure represents the dispersive analog of the
well-known steady spatial shock 
generated in the supersonic flow of a viscous compressible fluid past an
obstacle.  In this sense, it is the spatial counterpart of the
one-dimensional expanding dispersive shock \cite{kgk04}--\cite{hoefer}
generated in the evolution of large disturbances in a BEC.
In the simplest case, the nonlinear wave structure would consist of a
single spatial dark soliton given by a steady solution of the equations
governing the BEC flow.  Motivated by this physical consideration and
the results of experiments \cite{cornell05}, in this Letter we shall develop
the theory of spatial dark solitons in the framework of the
Gross-Pitaevskii (GP) mean field approach.

{\it Basic equations.}  The dynamics of a BEC is described to a good
approximation by the GP equation \cite{ps2003}
\begin{equation}\label{3-1}
   i\hbar\frac{\prt \psi}{\prt t}=-\frac{\hbar^2}{2m}\Delta\psi+
   V(\mathbf{r})\psi+g|\psi|^2\psi,
\end{equation}
where $\psi(\br)$ is the condensate order parameter and $g$ is an
effective coupling constant, with $g=4\pi\hbar^2a_s/m,$ $a_s$ being
the $s$-wave scattering length and $m$ the atomic mass. Here
$V(\mathbf{r})$ denotes the potential of the external forces acting
on the condensate, for example the confining potential of the trap
and/or the potential arising due to the presence of an obstacle
inside the BEC. When the ``obstacle'' is formed by a laser beam and
the flow occurs due to the free two-dimensional expansion of the BEC
the trap
potential should be set equal to zero for the free expansion of a
BEC, and far enough from the obstacle we can neglect the obstacle
potential as well. Also, we are interested in steady flows, that is,
we suppose that the parameters of the flow change on a timescale
much slower than the transient timescale for the establishment of
the wave pattern of interest.  To this end, we seek solutions of
Eq.~(\ref{3-1}) with $V(\mathbf{r})=0$ of the form
\begin{equation}\label{3-2}
   \psi(\br)=\sqrt{n(\br)}\exp\left(\frac{i}{\hbar}
   \int^{\br}{\bu}(\br')d\br'  \right ) \exp\left(-\frac{i\mu}{\hbar}t\right),
\end{equation}
where $n(\br)$ is the density of atoms in the BEC, ${\bf u}(\br)$
denotes its velocity field and $\mu$ is the chemical potential.
It is now convenient to introduce the dimensionless variables
$
   \tilde{\br}=\br/\sqrt{2}\xi,\,
   \tilde{n}=n/n_0,\, \tilde{\bf u}={\bf u}/c_s,
$ where $n_0$ is a characteristic density of atoms, equal to their density at
infinity, $\xi=\hbar/\sqrt{2mn_0g}$ is the healing
length and $c_s=\hbar/\sqrt{2}m\xi$ is the sound velocity in a BEC
of density $n_0$. Substituting Eq.~(\ref{3-2}) into
(\ref{3-1}) and separating real and imaginary parts we obtain a
system of equations for the density $n(x,y)$ and the two components of
the velocity field $\bu=(u(x,y),v(x,y))$,
\begin{equation}\label{3-3}
\begin{split}
   (nu)_x+(nv)_y=0,\\
   uu_x+vu_y+n_x+\left(\frac{n_x^2+n_y^2}{8n^2}-
   \frac{n_{xx}+n_{yy}}{4n}\right)_x=0,\\
   uv_x+vv_y+n_y+\left(\frac{n_x^2+n_y^2}{8n^2}-
   \frac{n_{xx}+n_{yy}}{4n}\right)_y=0,
   \end{split}
\end{equation}
where we have omitted tildes for convenience. If we restrict our
consideration to the vortices-free potential flows with vanishing
curl of the velocity field,
\begin{equation}\label{4-1}
    u_y-v_x=0,
\end{equation}
then the second and third equations in (\ref{3-3}) can be integrated
once to give a generalization of the Bernoulli theorem to dispersive
2D hydrodynamics:
\begin{equation}\label{4-2}
    \frac12(u^2+v^2)+n+\frac1{8n^2}(n_x^2+n_y^2)-
    \frac1{4n}(n_{xx}+n_{yy})=\mathrm{const}.
\end{equation}
 Equations (\ref{4-1}),
(\ref{4-2}) and the first equation of (\ref{3-3}) comprise the
system
governing the BEC potential flow.

Our aim now is to find the solution of this system under the
conditions that the BEC flow is uniform at infinity
\begin{equation}\label{3-4}
   n=1,\quad u=M,\quad v=0\quad\text{at}\quad |x|\to\infty,
\end{equation}
where $M$ denotes the ratio of the asymptotic velocity of the
flow to the sound velocity, i.e.\ the Mach number.

{\it Oblique dark soliton solution.} Let us seek a solution of the
form $n=n(\theta)$, $u=u(\theta)$, $v=v(\theta)$, where
$\theta=x-ay$ and $a$ denotes the slope of the soliton center
location with the $y$ axis. Substitution of this {\it ansatz\,} into
(\ref{4-1}) and the first equation of (\ref{3-3}), followed by a
simple integration, yields expressions for the velocity components
in terms of the density
\begin{equation}\label{4-3}
    u=\frac{M(1+a^2n)}{(1+a^2)n},\quad v=-\frac{aM(1-n)}{(1+a^2)n},
\end{equation}
where the integration constants were chosen according to
condition (\ref{3-4}). Then substitution of (\ref{4-3}) into
(\ref{4-2}), with a proper choice of the constant in the right-hand
side, leads to the equation
\begin{equation}\label{4-4}
    \tfrac14(1+a^2)(n_\theta^{2}-2nn_{\theta \theta})+2n^3-
    (2+p)n^2+p=0,
\end{equation}
where
\begin{equation}\label{4-5}
    p=M^2/(1+a^2).
\end{equation}
It is easily checked that Eq.~(\ref{4-4}) has the integral
\begin{equation}\label{4-6}
    \tfrac14(1+a^2)n_\theta^{2}=(1-n)^2(n-p)
\end{equation}
where, again, the integration constant is chosen in accordance
with condition (\ref{3-4}). Simple integration of this
equation finally yields the desired solution in the form of a dark
soliton for the density
\begin{equation}\label{5-1}
    n(\theta)=1-\frac{1-p}{\cosh^2[\sqrt{1-p}\,\theta/\sqrt{1+a^2}]}.
\end{equation}
The velocity components can then be found by substitution of this
solution into Eqs.~(\ref{4-3}). The inverse half-width of the
soliton in the $x$-direction is
\begin{equation}\label{5-2}
    \kappa=2\sqrt{\frac{1-p}{1+a^2}}.
\end{equation}
Thus formulae (\ref{5-1}) and (\ref{5-2}) give the exact dark
spatial soliton solution of the GP equation. We shall call it
``oblique'' because it is always inclined with respect to the
direction of the supersonic flow. Numerical solutions below show
that such solutions are stable for $M>1$.

{\it Small amplitude Korteweg-de Vries (KdV) limit.} As is clear
from (\ref{5-1}) the small amplitude limit is achieved when $1-p\ll
1$. Then the parameters $a$ and $p$ can be expressed in terms of
$\kappa$ and $M$ from Eqs.~(\ref{4-5}) and (\ref{5-2}) as
\begin{equation}\label{5-6}
    a\cong \sqrt{M^2-1}+\frac{M^4\kappa^2}{8\sqrt{M^2-1}},\quad
    1-p\cong\frac14\kappa^2M^2.
\end{equation}
The density profile in this limit becomes then familiar KdV soliton
\begin{equation}\label{5-5}
    n\cong 1-\frac{M^2\kappa^2}{4\cosh^2[\kappa(x-ay)/2]},
\end{equation}
where $\kappa\ll 1/M$.

Note that the slope $a=\sqrt{M^2-1}$ corresponds exactly to the
Cherenkov cone, so that in its vicinity the small amplitude solitons
are located inside it.
 This approximation corresponds to the KdV limit
of the potential-free GP equations (\ref{3-3}).
Indeed, if we assume
the series expansions
$(\eps\ll 1)$
\begin{equation}\label{6-1}
    n=1+\eps n_1+\ldots,\quad u=M+\eps u_1+\ldots,\quad
    v=\eps v_1+\ldots
\end{equation}
and introduce the scaling of the independent variables $
\zeta=\eps^{1/2}(x-ay),\quad \tau=\eps^{3/2}y$,  then standard
reductive perturbation theory leads to the KdV equation for the
density disturbance $n_1$
\begin{equation}\label{6-3}
    \partial_\tau n_{1}-\frac{3M^2}{2\sqrt{M^2-1}}n_1 \partial_\zeta n_{1}+
    \frac{M^4}{8\sqrt{M^2-1}}\partial^3_{\zeta\zeta\zeta}n_{1}=0,
\end{equation}
with the well-known soliton solution of this equation equivalent to
(\ref{5-5}), (\ref{5-6}).

{\it Nonlinear Schr\"odinger (NLS) equation limit.} Another
important limit corresponds to large slopes $a^2\gg 1$.  For this limit
we introduce the parameter $\la$ as $p\cong M^2/a^2=\la^2$, that is
$a=\pm M/\la$. The soliton solution (\ref{5-1}) can then be approximated as
\begin{equation}\label{6-4}
    n\cong1-\frac{1-\la^2}{\cosh^2[\sqrt{1-\la^2}(y\pm\la x/M)]}.
\end{equation}
This is exactly the solution $n=|\Psi|^2$ of the NLS equation
\begin{equation}\label{nls-1D}
   i\Psi_T+\Psi_{YY}-2|\Psi|^2\Psi=0
\end{equation}
for the complex variable $\Psi=\sqrt{n}\exp\left(i\int^Y
v(Y',t)dY'\right)$, where $T=x/2M$ and $Y=y$. This equation was
derived in \cite{ek-pla06} from the GP equations (\ref{3-3}) for the
highly supersonic ($M\gg1$) flow of a BEC past a slender body.


{\it Generation of oblique solitons in a BEC.} Let us now consider
the supersonic flow of a BEC past an obstacle.  If the obstacle is
small (e.g.\ an impurity), then linear sound waves are generated at
finite distances which form a Cherenkov cone \cite{AP04}. Large
obstacles generate spatial dispersive shocks which can be viewed as
trains of interacting dark spatial solitons inside the Cherenkov
cone. The theory of the generation of spatial dispersive shocks has
been developed in much detail for supersonic flows past a slender
body when such a flow can be described by the KdV equation
\cite{GKKE95}. Analogous theory for the NLS equation case was
developed in \cite{ek-pla06}. However, in real experiments the
obstacles cannot be considered as slender bodies and the flow is not
highly supersonic, hence fully nonlinear solutions of the GP
equation, such as Eq.~(\ref{5-1}), should be used for the
quantitative description of spatial dispersive shocks in a BEC. Here
we shall use numerical solutions of the GP equation to demonstrate
that the structures generated by an obstacle inserted into the
supersonic BEC flow indeed contain the oblique dark solitons given
by Eq.~(\ref{5-1}).

To make this process clearer, numerical solutions of the
time-dependent GP equation (\ref{3-1}), expressed in non-dimensional
variables as
\begin{equation}\label{7-1}
    i\psi_t=-\frac12(\psi_{xx}+\psi_{yy})+V(x,y)\psi+|\psi|^2,
\end{equation}
were studied, where $\widetilde{\psi}=\psi/\sqrt{n_0}$ and
$\tilde{t}=(gn_0/\hbar)t$, with the other variables defined as in
Eq.~(\ref{3-2}).  From now on, the tildes will be omitted.  The
potential $V(x,y)$ corresponds to the interaction of the condensate
with the obstacle.  Since the detailed behavior of the potential
should not be critical for the formation of solitons far from the
obstacle, it can be safely modeled by an impenetrable disk. In our
simulations its radius was set to $r=1$ in our non-dimensional
units.  Initially at $t=0$ there is no disturbance in the
condensate, so that it is described by the plane wave function
$\left.\psi(x,y)\right|_{t=0}=\exp(iMx)$ corresponding to a uniform
condensate flow.  To be specific, let us take $M=5$.  Several stages
of the numerically calculated BEC density evolution are shown in
Fig.~1.
\begin{figure}[bt]
\includegraphics[width=8cm,height=5cm,clip]{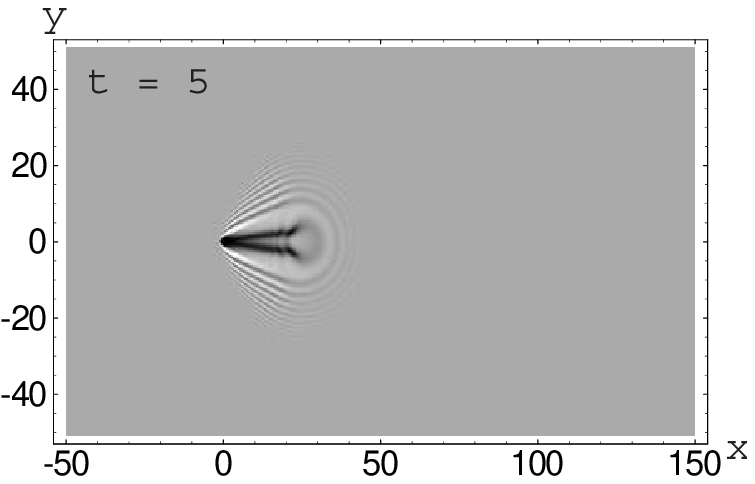}
\includegraphics[width=8cm,height=5cm,clip]{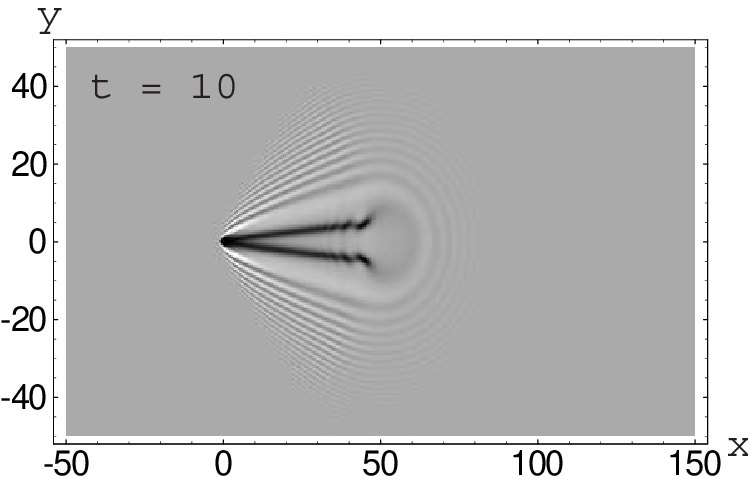}
\includegraphics[width=8cm,height=5cm,clip]{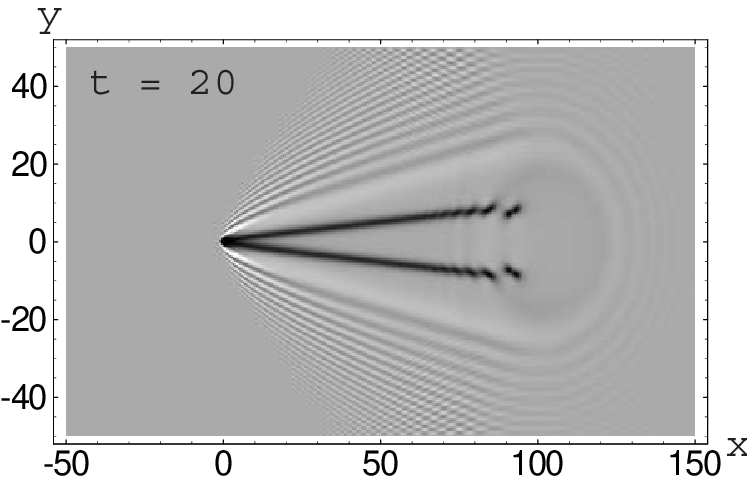}
\caption{Emergence of a pair of oblique dark solitons after
``switching on'' supersonic flow ($M=5$) past a disk-shaped
impenetrable obstacle of radius $r=1$ located at $(x=0,y=0)$.  The
direction of the flow is from left to right. Density plots are shown
for three times $t=5$, $t=10$ and $t=20$.  The dark structures
correspond to oblique dark solitons, which in turn generate the
vortex streets near the end points.} \label{fig1}
\end{figure}

It can be clearly seen how a pair of oblique solitons is gradually
formed behind the obstacle.  Their length grows with time and,
except in the vicinity of the end points, the density distributions
do not demonstrate any vorticity, which agrees with our assumption
of potential flow (\ref{4-1}). However, the flow cannot be
considered as stationary and potential near the end points.  This is
manifested by the presence of vortices behind the end points of the
spatial solitons.  One may interpret these vortices as a ``vortex
street'' \cite{WMCA} radiated by spatial dark solitons. However far
enough away from these vortex end points, the flow can be considered
stationary.  The establishment time of the stationary profile can be
estimated by the soliton width divided by the sound velocity, which
is $t\sim2$ in our solutions.  This is much less than the value
$t=20$ for the last plot of Fig.~1. The parameter $p$ from
Eq.~(\ref{4-5}) was calculated using the value of the slope $a$
inferred from the numerical solution. A comparison of the
theoretical profile of the oblique dark soliton given by
Eq.~(\ref{5-1}) with the corresponding part of the density profile
in the full numerical solution is shown in Fig.~2.  The excellent
agreement between these two profiles confirms that the line patterns
in Fig.~1 are indeed oblique dark solitons generated by the
obstacle. This agreement also justifies the assumptions made in the
derivation of the analytic solution (\ref{5-1}).  Note that, along
with the soliton, a small amplitude dispersive wave packet is
generated, which corresponds to the ``non-solitonic'' part of the
density perturbation induced by the obstacle.  This wave packet
spreads out as distance from the obstacle increases and eventually
fades away.
\begin{figure}[bt]
\includegraphics[width=8cm,height=5.5cm,clip]{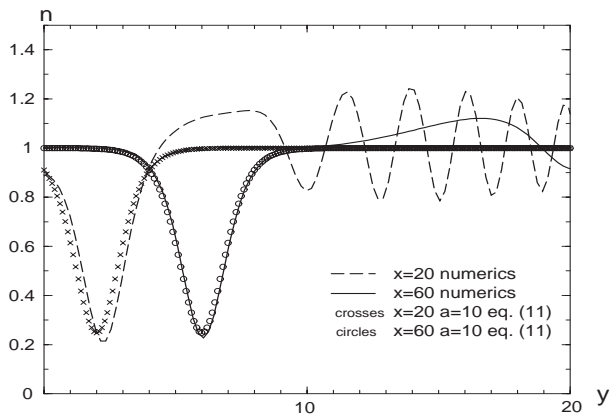}
\caption{Cross sections of the density distributions for $x=20$
(dashed line), $x=60$ (solid line) and $y>0$ obtained from numerical
solution of the GP equation (\ref{7-1}). These are compared with
soliton profiles (\ref{5-1}) with slope $a=10$ shown as functions of
$y$ at the same values of $x$ ($x=20$ corresponds to ``crosses'' and
$x=60$ to ``circles'').  } \label{fig2}
\end{figure}

In the above simulations, the parameters of the obstacle have been
chosen so that only a single soliton is generated at each side of
the obstacle.  However, with the increase of the size $r$ of the
obstacle the number of solitons is also expected to increase. This
effect is demonstrated in Fig.~3 in which two symmetric fans of
solitons can be seen behind the obstacle.  As expected, the depth of
the dark solitons grows with the increase of the slope $a$ with
respect to the $y$ axis. Thus, the oblique dark solitons can be
viewed as ``building blocks'' in more complicated patterns arising
in supersonic flows of a BEC. This figure also demonstrates the
robustness of the phenomenon for different obstacle parameters. The
whole structure in Fig.~3 represents a pair of dispersive shocks
generated in the supersonic flow of a BEC past an obstacle.  Such
shocks were considered in \cite{ek-pla06} in the limiting case of
highly supersonic flow ($M\gg1$) and slender obstacles ($a\gg1$).
Our present numerical simulations show that spatial dispersive
shocks represent a general phenomenon caused by the interplay of
dispersive and nonlinear effects described by the full GP equation.
We speculate that it can be also be observed in flows of a BEC past
3D obstacles.
\begin{figure}[bt]
\includegraphics[width=8cm,height=5.5cm,clip]{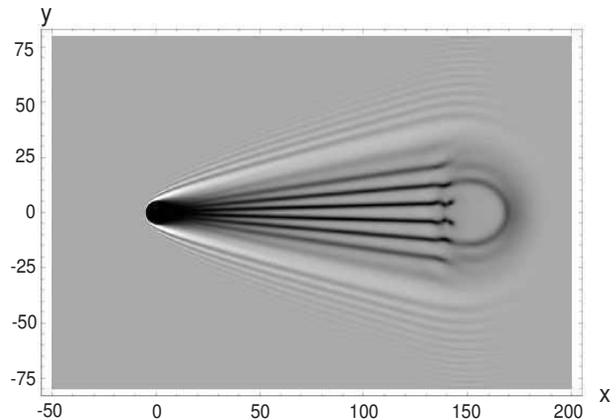}
\caption{Density plot at the moment $t=30$ for the supersonic flow
($M=5$) past a disk-shaped impenetrable obstacle with radius $r=5$
located at $(x=0,y=0)$. } \label{fig3}
\end{figure}

The above theory is based on the supposition that the flow of a BEC
is homogeneous and uniform enough for periods of time sufficient for
generation of oblique dark solitons. Let us indicate here the
physical conditions which should be satisfied for obtaining such a
flow in experiment. To be definite, we consider the example of 2D
flow in ``pancake'' geometry, that is the condensate is supposed to
be confined in axial direction by a strong harmonic potential. Let
the transversal frequency $\omega_\bot$ of the potential be small
enough so that the (radial) Thomas-Fermi density distribution is
applicable. We consider expansion of the BEC after switching off the
potential (see, e.g.~\cite{kam04}). The asymptotic as $t\gg
1/\omega_\bot$ solution of the cylindric GP equation for $r\ll
l\omega_\bot t$ where $l$ is the radius of the BEC before its
release from the trap, is given by $n(r,t)\cong
{\hbox{constant}}/({\omega_\bot t})^2$, $ u(r,t)\cong{r}/t$, that
is the density is practically uniform but varying with time. In the
same approximation, the local healing length $\xi$ and the local
sound velocity $c_s$ are given by $ \xi={\hbar t}/{ml},$
$c_s={l}/({\sqrt{2}t})$.  Hence, the local Mach number $ M \cong
{\sqrt{2}r}/{l}. $ Thus, we get a supersonic flow past obstacle if
we place it at the distance $d > l/\sqrt{2}$ from the axis.  Now,
the flow can be considered as uniform if the size of the obstacle
(scaled as healing length $\xi$ for a chosen moment of observation)
satisfies the condition $\xi/d\ll 1$. For $d\sim l$ this gives $
t\ll{ml^2}/{\hbar}$, i.e the flow past an obstacle is asymptotically
uniform for $\omega_{\bot} ^{-1} \ll t\ll{ml^2}/{\hbar}$. At last,
the characteristic time of establishing the soliton distribution, $
{\xi}/{c_s}\sim{m\xi^2}/{\hbar} $, obviously satisfies the above
inequality since $\xi\ll l$, so that the flow can be considered as
quasi-stationary. At the same time, our numerical simulations show that
oblique solitons are generated even in non-uniform and nonhomogeneous
flows past obstacles, that is the phenomenon is very robust with respect
to change of parameters of the flow.

To summarize, we have found an exact oblique dark soliton solution
of the GP equation and demonstrated numerically that such solitons
can be generated by obstacles inserted into the supersonic flow of a
BEC.

AMK thanks EPSRC and RFBR (grant 05-02-17351) and
AG thanks FAPESP and CNPq for financial support.

\end{document}